\documentclass[12pt,british,a4paper]{iopart}
\usepackage{babel}%,amsmath,amssymb}
\usepackage{iopams}
\usepackage{graphicx}
\usepackage{cite}

\usepackage{bm}
\usepackage[sans]{dsfont}
\usepackage[mathscr]{euscript}
\usepackage[latin1]{inputenc}
\usepackage{geometry}
\usepackage{epstopdf}

\newcommand{\eqref}[1]{(\ref{#1})}

\begin{document}

%\title{Approximation of an evolution operator with the help of projection operator technique}
\title{Projection operator based expansion of the evolution operator}

%\author{Vitalii Semin$^{1,2}$ and Francesco Petruccione$^{1,2}$}

\author{Vitalii Semin}

\address{Samara National Research University, 34, Moskovskoe shosse, Samara, 443086, Russian Federation} \ead{semin@ssau.ru}

\author{Francesco Petruccione}

\address{Quantum Research Group, School of Physics and National
Institute for Theoretical Physics, University of KwaZulu-Natal,
Durban, 4001, South Africa} \ead{petruccione@ukzn.ac.za}

%\affiliation{{$^1$Quantum Research Group, School of Chemistry and Physics,
 %University of KwaZulu-Natal, Durban, 4001, South Africa},
% {$^2$National Institute for Theoretical Physics (NITheP), KwaZulu-Natal, South Africa}}

\date{\today}

\begin{abstract}
The not necessarily unitary evolution operator of a finite dimensional quantum system is studied with the help of a projection operators technique. Applying this approach to the Schr\"{o}dinger equation allows deriving an alternative expression for the evolution operator, which differs from the traditional chronological exponent. An appropriate choice of the projection operators results in the possibility to study the diagonal and non-diagonal elements of the evolution operator separately. The suggested expression implies a particular form of perturbation expansion, which leads to a new formula for the short time dynamics. The new kind of perturbation expansion can be used to improve the accuracy of the usual chronological exponent significantly. The evolution operator for any arbitrary time can be efficiently recovered using the semigroup properties. The method is illustrated by two examples, namely, the dynamics of a three-level system in two nonresonant laser fields and the calculation of the partition function of a finite XY-spin chain.
\end{abstract}


\maketitle

\section{Introduction.}
%Quantum mechanics is a keystone of the modern science. The variety of phenomena of the surrounding world are explained  by quantum mechanics. Particularly quantum mechanics explains such amazing phenomena as superconductivity \cite{SC}, superfluidity \cite{SF}, superradiation \cite{DIKE} and even photosyntheses \cite{PHOTO}.

%Dynamics of non-relativistic quantum systems is completely defined by the Schr\"{o}dinger equation for a wave vector or the Liouville-von Neumann equation for a density operator. An alternative approach is based on the Heisenberg equation for observables. The results of any above mentioned equations can be reproduced through the action of only one operator, namely, the so-called evolution operator, which satisfies the Schr\"{o}dinger evolution equation. In this sense 
The evolution operator is the most fundamental object in quantum mechanics, in the sense that the time dynamics of any  quantum object can be defined through the action of the evolution operator.
Equations for the evolution operator often represent a system of linear differential equations. In general, such equations cannot be solved analytically, and are usually studied with the help of approximations. %This is a reason to introduce some  approximation techniques.
Most approaches are actually some modification of perturbation expansion for the evolution operator, which describe only the short time behaviour. Typical examples are the Dyson series \cite{DYSON}, different variants of geometrical integrators, such as Magnus \cite{MAG, MAG2}, Fer or Wilcox \cite{MAGNUS} methods, the exponential splitting method \cite{SPLIT}.
The long-time dynamics is traditionally recovered with the help of the Lie-Trotter formula \cite{LTK} or its analogues \cite{CHERNOFF}, or, alternatively, the Feynman path integrals \cite{FEYNMAN}.

An alternative way to study the evolution operator is direct numerical integration. There is a vast number of different numerical algorithms \cite{num1,num2}, which may be used for an investigation of the evolution operator. In many applications, the direct numerical integration using high-performance algorithms is preferable than the perturbation approach. Nevertheless, in cases when one is interested only in some part of the evolution operator, for example, in the calculation of a partition function, which is directly connected with the evolution operator due to the Wick rotation \cite{WICK}, the numerical methods do not give much advantage. Especially, this problem is crucial in higher dimensions. It is evident that none of the above-mentioned methods allows extracting necessary elements without calculation of the full evolution operator.

In this paper, we study the evolution operator with the help of projection operators. Using the projection operators we split the Schr\"{o}dinger equation for the evolution operator into two parts and study each part separately. The study of these parts is performed with the help of a procedure which is an analogue of the important time-convolutionless technique in the theory of open quantum systems \cite{toqs}. Moving to the simplest possible form of the method we concretise the projection operator, which extracts the diagonal part of the evolution operator. In this case, the system for the diagonal elements decomposes into a set of independent linear differential equations and can be solved analytically. The full evolution operator is represented as a product of two matrices one of which is diagonal and can be studied separately. We consider this expression as a starting point for the development of a particular form of perturbation theory. Interestingly, the new perturbation expansion allows improving the results following from the usual chronological exponent of the corresponding order. The perturbation expansion up to the second order approximates the short time dynamics of the evolution operator. This approximation can be used for recovering the evolution operator using the semigroup property of the evolution operator iteratively. The technique is illustrated with the examples of the dynamics of a three-level system in two non-resonant laser fields and the calculation of the partition function for the finite XY-spin chain.

The paper is organised as follow. In Sec. II we discuss the general properties of the evolution operator. Sec. III deals with the projection operator technique in the context of the evolution operator problem. In this section we derive in details our most important result Eq. (\ref{new}) and, moreover, discuss the new form of perturbation expansion. We qualitatively compare different forms of perturbation expansions with the suggested one. The suggested formalism is applied to two concrete physical systems in Sec. IV. We conclude in Sec. V.

\section{Evolution operator.}
%In this paragraph we describe the main properties of the evolution operator, which will be used further. 
The evolution operator satisfies the Schr\"{o}dinger evolution equation
\begin{equation}\label{evol}
\dot{U}=-i HU,
\end{equation}
with the initial condition $U(t_0,t_0)=I,$ where $H$ is the system Hamiltonian, which is not necessarily hermitian,  and $I$ is the identity operator (we set $\hbar=1$). The formal solution of the Schr\"{o}dinger equation can be written in the form of the so-called chronological exponent or time-ordered product
\begin{eqnarray}\label{chron}
U(t,t_0)=\exp_-\left[-i\int_{t_0}^t H(s)ds \right]\\=I+\sum_{n=1}^\infty(-i)^n\int_{t_0}^tds_1\int_{t_0}^{s_1}ds_2\cdots\int_{t_0}^{s_{n-1}}ds_nH(s_1)H(s_2)\cdots H(s_n).\nonumber
\end{eqnarray}
%The chronological exponent \eqref{chron} is often calculated perturbatively. The perturbation expansion is a base of a broad class of approximation techniques \cite{LTK, CHERNOFF, FEYNMAN}.
Also, we need an inverse evolution operator. Note, that the evolution operator may not be unitary for non-hermitian Hamiltonians. The inverse evolution operator is governed by the following equation
\begin{equation}\label{inverse}
\dot{U}^{-1}=iU^{-1} H.
\end{equation}
The solution of the above equation can be expressed through the antichronological exponent
\begin{eqnarray}\label{antichron}
U^{-1}(t,t_0)=\exp_+\left[i\int_{t_0}^t H(s)ds \right]\\=I+\sum_{n=1}^\infty(i)^n\int_{t_0}^tds_1\int_{t_0}^{s_1}ds_2\cdots\int_{t_0}^{s_{n-1}}ds_nH(s_n)H(s_{n-1})\cdots H(s_1).\nonumber
\end{eqnarray}
The evolution operator possesses the semi-group property
\begin{equation}\label{semigroup}
U(t,t_0)=U(t,s)U(s,t_0),
\end{equation} 
which is the basis of different approximation techniques, such as the Lie-Trotter-Kato formula \cite{LTK} and the Chernoff theorem \cite{CHERNOFF}. 
%The main idea of such techniques is using the semigroup property  \eqref{semigroup} to devide the time interval $t-t_0$ into large number subintervals and 

%The majority of such techniques are based on Lie-Trotter formula and its generalizations. 
% The Lie-Trotter formula can be expressed as 
%\begin{equation}
%\end{equation}

\section{Projection Operator Techniques.} %In this paragraph we derive an expression, which is the basis of our approximation technique. 
Let us introduce the pair of projection operators $\mathcal{P}=\mathcal{P}^2$ and $\mathcal{Q}=\mathcal{Q}^2$ with the following properties: i)  $ \mathcal{P}\mathcal{Q}=\mathcal{Q}\mathcal{P}=0;$ ii) $\mathcal{P}+\mathcal{Q}=I;$ and iii) $\frac{\partial}{\partial t}\mathcal{P}=\mathcal{P}\frac{\partial}{\partial t}.$ In the above expressions $I$ is the identity operator and $t$ is time. A specific form for projection operator will be discussed later. Now, we let the projection operators act on both sides of the Schr\"{o}dinger equation (\ref{evol}) to derive 
\begin{eqnarray}
\mathcal{P}\dot{U} =-i\mathcal{P}H(\mathcal{P}+\mathcal{Q})U, \label{proj1}\\
\mathcal{Q}\dot{U} =-i\mathcal{Q}H(\mathcal{P}+\mathcal{Q})U. \label{proj2}
\end{eqnarray}
The formal solution of Eq.~(\ref{proj2}) is
\begin{equation}\label{irrel}
\mathcal{Q}U=\mathcal{G}(t,t_0)\mathcal{Q}I-i\int_{t_0}^t ds \mathcal{G}(t,s)\mathcal{Q}H\mathcal{P}U(s,t_0),
\end{equation}
where we introduce $\mathcal{G}(t,s)=\exp_-\left\{-i\int_s^t dt'\mathcal{Q}H(t')\right\}.$ %and $\exp_\mp$.

Substituting (\ref{irrel}) for (\ref{proj1}) leads to the integro-differential equation of the form
\begin{equation}
\mathcal{P}\dot{U} =-i\mathcal{P}H\mathcal{P}U-i\mathcal{P}H\left(\mathcal{G}(t,t_0)\mathcal{Q}I-i\int_{t_0}^t ds \mathcal{G}(t,s)\mathcal{Q}H\mathcal{P}U(s,t_0) \right).\label{rel}\end{equation}

The evolution operator is defined by the solution of Eqs.(\ref{irrel})-(\ref{rel}) and has the obvious form, namely, $U=\mathcal{P}U+\mathcal{Q}U.$ The technique used in deriving Eqs.\eqref{irrel}-\eqref{rel} is  analogue to the famous Nakajima-Zwanzig projection operator technique in the theory of open quantum systems \cite{toqs, NAKAJIMA, ZWANZIG}. Notice, that Eqs. (\ref{irrel})-(\ref{rel}) are absolutely equivalent to the initial Schr\"{o}dinger equation (\ref{evol}). At the same time, the specific choice of projection operators may simplify an investigation of the evolution operator. An obvious advantage of the approach is that the  Eqs. (\ref{irrel})-(\ref{rel}) allow to reduce the number of equations in the system (\ref{evol}).
%develop a specific form of perturbation theory, which may be useful.

Because the above theory, in fact, does not lead to a new perturbation expansion let us continue  the formal transformation as it is usually done in the time-convolutionless projection operator technique \cite{toqs}. We substitute the identity $U(s,t_0)=U^{-1}(t,s)(\mathcal{P}+\mathcal{Q})U(t,t_0),$ where the inverse operator is understood as an antichronological exponent (\ref{antichron}), into Eq.~(\ref{irrel}) and after some algebra we get
\begin{equation}\label{irrel2}
\mathcal{Q}U(t,t_0)=[I-\Sigma(t)]^{-1}(\Sigma(t)\mathcal{P}U(t,t_0)+\mathcal{G}(t,t_0)\mathcal{Q}I),
\end{equation}
where we introduced the superoperator $\Sigma(t)=-i\int_{t_0}^t ds \mathcal{G}(t,s)\mathcal{Q}H(s)\mathcal{P}U^{-1}(t,s).$
From Eq.~(\ref{proj1}) and Eq.~(\ref{irrel2}) we finally find
\begin{equation}\label{rel2}
\mathcal{P}\dot{U}(t,t_0)=\mathcal{K}(t)\mathcal{P}U(t,t_0)+\mathcal{I}(t)\mathcal{Q}I.
\end{equation}
In the above equation $\mathcal{K}(t)=-i \mathcal{P}H(t)[I-\Sigma(t)]^{-1}\mathcal{P}$ and $\mathcal{I}(t)=-i \mathcal{P}H(t)[I-\Sigma(t)]^{-1}\mathcal{G}(t,t_0)\mathcal{Q}.$
All the above expressions are exact and valid for any projection operator. Notice, that Eq. ~(\ref{irrel}),  (\ref{rel}), \eqref{irrel2} , \eqref{rel2} can be used to obtain an alternative expression for the evolution operator.

%o alternative calculation of  the evolution operator.

%On the other hand Eq.~\eqref{irrel2} and \eqref{rel2} have rather formal interest, because it is much easier to solve the Schr\"{o}dinger equation \eqref{evol} then find an exact form of the superoperators in Eq.~\eqref{irrel2} and \eqref{rel2}. None the less, Eq.~\eqref{irrel2} and \eqref{rel2} can be consider as a starting point for a systematic perturbation study of the evolution operator.

\subsection{Specific form  of projection operators and alternative expression for the evolution operator.}
We choose a specific form of projection operators and study Eq.~\eqref{irrel2} and \eqref{rel2} more precisely. Firstly, we divide the evolution operator into two parts, namely, the diagonal part and the  non-diagonal part. The projection operator which extracts the diagonal part has the following form
\begin{equation}\label{specproj}
\mathcal{P}A=\sum_i^n\mathrm{tr}(A E_{ii})E_{ii},
\end{equation}
and the additional projection operator is
\begin{equation}\label{specproj2}
\mathcal{Q}A=(I-\mathcal{P})A,
\end{equation}
where $E_{ij}$ is a $n\times n$ matrix with only one unit in the intersection of $i$th row and $j$th column and 0 elsewhere.

Such a choice of the projection operators allows to drop off terms proportional to $\mathcal{Q}I$ in Eqs. \eqref{irrel2} and \eqref{rel2}, because of they are equal to zero.

Secondly, we transform the system Hamiltonian to get rid of the diagonal elements
\begin{equation}\label{ham}
H_I(t)=\exp[i \int_{t_0}^t\mathcal{P}H(s)ds]\mathcal{Q}H(t)\exp[-i \int_{t_0}^t\mathcal{P}H(s)ds].
\end{equation}
The above transformation is equivalent to transition to the interaction picture with respect to the Hamiltonian $H_0(t)=\mathcal{P}H(t).$
%Now, it is clear that 
The total evolution operator is a product $U(t,t_0)=\exp[-i \int_{t_0}^t\mathcal{P}H(s)ds]\exp_-[-i \int_{t_0}^t H_I(s)ds]=U_0(t,t_0)U_I(t,t_0),$ where $U_I(t,t_0)$ satisfies the Schr\"{o}dinger equation \eqref{evol} with the Hamiltonian $H_I(t),$ and $U_0(t,t_0)$ satisfies the Schr\"{o}dinger equation \eqref{evol} with the Hamiltonian $H_0(t).$  This can be proved by the direct substitution of the total evolution operator in the Schr\"{o}dinger equation \eqref{evol}. Notice, that the operator $U_0(t,t_0)=\exp[-i \int_{t_0}^t\mathcal{P}H(s)ds])$  can always be  found in the  exact analytical form.  %The fundamental interest offers the non-diagonal evolution operator $U_I(t,t_0)$ and below we focus specifically on this operator.
Now it can be checked that for the  projection operators \eqref{specproj} and  \eqref{specproj2} the following relations hold $\mathcal{P}H_I(t)\mathcal{P}=0,$ $\mathcal{Q}I=0,$ and  $\mathcal{P}I=I.$ 

Below we will show that the transformation \eqref{ham} allows to reduce number of terms in the perturbation expansions of the operator $U_I(t,t_0)$ and, thus, simplify the study of a quantum systems with the help of the suggested method.  

Eq.~ \eqref{rel2} with the projection operators \eqref{specproj} and \eqref{specproj2} actually represents a homogeneous system of $n$ uncoupled differential equations of the first order. Such a system can be easily solved in quadratures and the solution has the form $\mathcal{P}U_I(t,t_0)=\exp\left[\int_{t_0}^t \mathcal{K}(s)ds\right]I$. The solution of Eq.~ \eqref{rel2} after substitution for \eqref{irrel2} gives the following expression for the full evolution operator $U_I(t,t_0)=\mathcal{P}U_I(t,t_0)+\mathcal{Q}U_I(t,t_0)$ 
\begin{eqnarray}\label{new}
\!\!\!\!\!\!\!\!\!\!\!\!U_I(t,t_0)=\exp\left[\int_{t_0}^t \mathcal{K}(s)ds\right]I+[1-\Sigma(t)]^{-1}\Sigma(t)\exp\left[\int_{t_0}^t \mathcal{K}(s)ds\right]I\\=\left[I-\Sigma(t)\right]^{-1}\exp\left[\int_{t_0}^t \mathcal{K}(s)ds\right]I.\nonumber  
\end{eqnarray}
The expression \eqref{new} is the main result of this paper. Notice, that the right-hand side of Eq.~\eqref{new} is the superoperator which acts on the identity operator. At the same time, the exponent is actually a diagonal matrix and the superoperator $\left[I-\Sigma(t)\right]^{-1}$ acts trivially on this matrix. Thus, the expression \eqref{new} represents the product of two matrices, namely the diagonal matrix $\left( \exp\left[\int_{t_0}^tds \mathcal{K}(s)ds\right]I\right) $ and the matrix $\left( \left[I-\Sigma(t)\right]^{-1}I\right).$ In other words one can calculate the two matrices independently and than multiply them. At the same time, the exact calculation of the rigth-hand side of Eq.~\eqref{new} already implies the knowledge of the inverse evolution operator. Moreover, even if the inverse operator is known this calculation is not a trivial task. Nevertheless, Eq.~\eqref{new} is the basis for an alternative form of the perturbation expansion, which we will develop in the next section.

\subsection{Perturbation expansion for the evolution operator.}
Suppose that the Hamiltonnian in \eqref{evol} depends on some perturbation parameter $\alpha.$  From the explicit form of the superoperator  $\mathcal{K}(t)=-i \mathcal{P}H(t)[I-\Sigma(t)]^{-1}\mathcal{P}$ it is obvious that the perturbation expansion is completely
 determined by the expansion of $[I-\Sigma(t)]^{-1}=\sum_{i=0}^\infty \Sigma(t)^i=\sum_{i=0}^\infty( \sum_{j=1}^\infty\alpha^j \Sigma_j(t))^i \approx I+\alpha\Sigma_1(t)+\alpha^2(\Sigma_1(t)^2+\Sigma_2(t))+O(\alpha^3).$ Here we suppose that the superoperator $[1-\Sigma]^{-1}$ exists and can be expanded in a geometric series.
 Notice, that the superoperator $\Sigma(t)$ consists both of the chronological exponent  $\mathcal{G}(t,s)=\exp_-\left\{-i\int_s^t dt'\mathcal{Q}H(t')\right\},$ and of the  antichronological exponent  $U^{-1}(t,t_0)=\exp_+\left[i\int_{t_0}^t H(s)ds \right].$ Thus, the perturbation expansion has no well defined time-ordering.
 Using the explicit expression for $\Sigma(t)$  one can find that 
\begin{equation}\label{sig1}
\Sigma_1(t)=-i\int_{t_0}^t \mathcal{Q}H(s) \mathcal{P}ds,
\end{equation}
 which immediately gives $\Sigma_1(t)^2=0,$ due to the identity $\mathcal{Q}\mathcal{P}=0,$ and 
\begin{eqnarray}\label{sig2}
\Sigma_2(t)=-\int_{t_0}^t dt_1\int_{t_0}^{t_1}dt_2\left(\mathcal{Q}H(t_1) \mathcal{Q}H(t_2)\mathcal{P}-\mathcal{Q}H(t_2) \mathcal{P}H(t_1)\right).
\end{eqnarray}
 
Now, to simplify the second order term we may get rid of the diagonal elements of the Hamiltonian, by using the transformation \eqref{ham}. In other words, we transform the Hamiltonian to the interacting picture. Notice, that this transformation simplifies the method. In the interaction picture the identity $\mathcal{P}H_I(s) \mathcal{P}=0$ can be easily proven and the second term in Eq. \eqref{sig2} can be put equal to zero. The resulting expressions for the superoperator $\mathcal{K}(t)$ up to the second order is 
\begin{eqnarray}
\mathcal{K}(t)=-\int_{t_0}^tds\mathcal{P}H_I(t)H_I(s)\mathcal{P}+O(\alpha^3),\label{1p}\\
\left[ I-\Sigma(t)\right] ^{-1}=I-i\int_{t_0}^tds\mathcal{Q}H_I(s)\mathcal{P}\label{2p}\\
-\int_{t_0}^tds\int_{t_0}^sds_1\mathcal{Q}H_I(s)H_I(s_1)\mathcal{P}+O(\alpha^3).\nonumber
\end{eqnarray}

The general term of the $n$th order may be found using a method similar to the cumulant expansion for stochastic differential equations suggested by van Kampen in \cite{KAMPEN}. The resulting expression for $\mathcal{K}_n(t)$ has the following form

\begin{eqnarray}
\mathcal{K}_n(t)=(-i)^n\int_{t_0}^t dt_1\int_{t_0}^{t_1} dt_2...\int_{t_0}^{t_{n-2} }dt_{n-1}\label{full}\\
\times\sum (-1)^q\mathcal{P}H(t) ...H(t_i)\mathcal{P}H(t_j)...H(t_k)\mathcal{P}H(t_l)...H(t_m)\mathcal{P}...\mathcal{P},\nonumber
\end{eqnarray}
where the right-hand side is defined as follow \cite{toqs}. First, one writes down a string of the form $\mathcal{P}H ... H\mathcal{P}$ with $n$ factors of $H$ in between two $\mathcal{P}$s. Next one inserts an arbitrary number $q$ of factors $\mathcal{P}$ between the $H$s such that at least one $H$ stands between two successive $\mathcal{P}$ factors. The resulting expression is multiplied by a factor $(-1)^q$ and all $H$s are furnished with a time argument: The first one is always $H(t)$. The remaining $H$s carry any permutation of the time arguments $t_1, \, t_2,\, ...\, t_{n-1}$ with the only restriction that the time arguments in between two successive $\mathcal{P}$s must be ordered chronologically. Finally, the resulting expression is obtained by a summation over all possible insertions of $\mathcal{P}$  factors and over all allowed distributions of the time arguments.

The same procedure can be used to obtain the expansion of $(I-\Sigma(t))^{-1},$ if we  remember that $\mathcal{K}(t)=-i\mathcal{P}H(t)(I-\Sigma(t))^{-1}\mathcal{P},$ i.e., $\mathcal{K}_n(t)=-i\mathcal{P}H(t)\Sigma_{n-1}(t)\mathcal{P}.$

The above perturbation expansions is alternative to the Dyson series. Nevertheless, the difference between the two expansions can be understood from the explicit form of the supeorperators, which consists of both chronological and anti-chronological exponents. Thus, the expansion of Eq. \eqref{new} does not have well defined chronological ordering, oppositely to  the Dyson series. Below we compare the two expansions for concrete examples.

%Notice, that expansion of any order of Eq. \eqref{new} does not preserve the properties of the exact evolution operator, such as unitarity. This fact distinguishes the suggested technique from the geometrical integrators methods, such as Magnus or Fer methods \cite{MAGNUS}. These  geometrical integrators involve operator exponents, which are quite difficult to calculate, and are usually treated approximately. The perturbation expansions \eqref{1p}-\eqref{2p} do not have such problems and can be used to recover the evolution operator with any accuracy, as it is discussed below.

\subsection{Scheme of recovery of the evolution operator.}
The Eq.~\eqref{new} with superoperators \eqref{1p}-\eqref{2p} reproduces the short time form of the evolution operator $U(t,t_0),$ where $\alpha(t-t_0)\ll 1$. To recover the evolution operator for any time we act in the spirit of the famous Lie-Trotter theory, namely, divide the time interval [0,t] into $N$ subintervals. Using the semigroup property of the evolution operator \eqref{semigroup} we can write
\begin{equation}\label{trot}
U(t,t_0)=U(t=t_N,t_{N-1})U(t_{N-1},t_{N-2})\dots U(t_1,t_0).
\end{equation}
Each operator $U(t_i,t_{i-1})$ is defined with the help of Eqs.~\eqref{new},\eqref{1p} and \eqref{2p}. Now, we indicate an interesting iterative expression, which allows to speed up the calculation
\begin{equation}\label{iter}
U(t,0)=U(t=t_N,t_{N-1})U(t_{N-1},0).
\end{equation}
The iterative scheme requires only one matrix multiplication per iteration. Moreover, the $U(t_N,t_{N-1})$ is a continuous operator on the time interval $[t_N,t_{N-1}].$ %defined by  \eqref{new}. 
Thus, the iterative scheme \eqref{iter} gives a continuous expression for the evolution operator.

%In the end of this section let us note the following. 
The Eq.~\eqref{new} is an identity and formally fulfilled for infinite dimensional systems. %(of course if $[I-\Sigma(t)]^{-1}$ exists). 
At the same time, calculation of the evolution operator for high dimensional systems with this expression is extremely difficult. Nevertheless, the same remark is correct for calculation of the usual truncated Dyson series or other approximation techniques.

\subsection{Comparison with other expansions}
There exists several different types of expansions which are used in various applications. The Dyson series \cite{DYSON} is probably the most typical one, and is just a time ordered product 
\begin{equation} \label{dyson}
U(t,t_0)=\exp_-\left[-i\int_{t_0}^t H(t')dt'\right].
\end{equation}
It can be easily proved that the Nakajima-Zwanzig projection operator technique leads exactly to the Dyson series for the full evolution operator independently of the concrete form of a projection operator.  In contrast, local in time forms of the projection technique consist of both chronological and anti-chronological exponents and, thus, have no well defined time ordering. This fact makes Eq. \eqref{new} absolutely different from the Dyson series, where the time ordering is well defined. This can already be  seen from the second order perturbation term \eqref{sig2}.

Another common type of expansion is  the Magnus series \cite{MAGNUS}. The Magnus series has the following general form
\begin{equation}\label{magnus}
U(t,t_0)=\exp[M(t,t_0)],
\end{equation}
where $M(t,t_0)=\sum_n^\infty M_n(t,t_0)\approx -i\int_{t_0}^t H(s)ds-\int_{t_0}^t ds_1\int_{t_0}^{s_1}ds[H(s_1),H(s)]+\cdots.$  The explicit form of the general $W_n$ can be found in \cite{MAGNUS}. The expansion \eqref{magnus} has a big advantage, namely, it preserves the main characteristic of the exact solution $U(t,t_0),$ for example unitarity, in any order of the perturbation expansion. None of the expansions \eqref{new} or \eqref{dyson} has such a property. It is clear that Magnus expansion does not coincide with the expansion \eqref{new} in any perturbation order. Similarly, Eq. \eqref{new} differs from any others type of geometrical integrators, such as Fer or Wilcox methods \cite{MAGNUS}.
Nevertheless, notice that the explicit calculation of the operator exponent in \eqref{magnus} (and similar methods) is associated with huge difficulties and can be done only approximately in most cases. This fact nullifies any advantages of the method in concrete applications. Both the expansions \eqref{new} and \eqref{dyson} don't have such a problem.

%One may try to connect Eq. \eqref{15} and the Magnus expansion using the expression $\sum_{n=1}\Omega_n(t)=\log [I-\Sigma(t)]^{-1}\exp\left[\int_{t_0}^t dt' \mathcal{K}(t')\right]I\approx  \log ([I-\Sigma(t)]^{-1}I)+R(t)+1/2[\log ([I-\Sigma(t)]^{-1}I),R(t)]_-+...,$ where $R(t)=\int_{t_0}^t dt' \mathcal{K}(t')I,$ and $[a,b]_-$ is the commutator of two operators.  To derive this expansion we used the Baker-Hausdorff formula. Sorting equal powers from the RHS and the LHS one gets the sought-for connection. The first two orders have the following form
%\begin{eqnarray}
%\Omega_1(t)=\Sigma_1(t)I+R_1(t)I=-i\int_{t_0}^t H(t')dt',\\
%\Omega_2(t)=\Sigma_2(t)I+R_2(t)+1/2[\Sigma_1(t)I,R_1(t)]_-.
%\end{eqnarray}
% where $R_i(t)=\int_{t_0}^tdt'K_i(t)I.$
 
%It seems that the explicit connection between Magnus series and Eq. \eqref{15} do not coincide with the known Magnus iterators \cite{MAGNUS} and lead to alternative form of the expansion. Also, the iterators of the second and higher order are not antihermitian for hermitian hamiltonians. So, the resulting expression will not have the main property of the geometrical integrators, namely, preserving of main characteristics of the exact solution.

%It is clear that any truncated expansions can reproduce only short time dynamics. The long-time dynamics may be reconstructed with the help of semi-group properties as discussed  in the previous section.

\section{Examples.}

\subsection{Three level $\Lambda$-system in two non-resonant laser fields.}
The interaction Hamiltonian of the model in the interaction pictures is 
\begin{equation}\label{ham1}
H(t)=\Omega_1(t)E_{32}+\Omega_2(t)E_{31}+\mathrm{h.c},
\end{equation}
where the $E_{ij}$ have been introduced earlier, $\Omega_i(t)=\Omega_i\exp[i\omega_i t],$ and $\Omega_i,\, \omega_i,$ are, respectively, the Rabi frequency of the $i$th external field and the detuning of the laser field frequency from the atomic transition frequency.

The direct calculation of Eq.~\eqref{1p}-\eqref{2p} gives
\begin{eqnarray}
\exp[-\int_{t_0}^t\mathcal{K}(t)dt]=\exp[E_{33}(f^*_1+f^*_2)+E_{22} f_1+E_{11}f_2]\\
\left[ I-\Sigma(t)\right] ^{-1}=I-i\int_{t_0}^t ds H(s)-\Omega_1\Omega_2(gE_{21}+\mathrm{h.c.}),
\end{eqnarray}
where $f_i(t,t_0)=-\Omega_i^2\int_{t_0}^tdt'\int_{t_0}^{t'} ds\exp[i(s-t')\omega_i]$ and $g(t,t_0)=\int_{t_0}^tdt_1\int_{t_0}^{t_1}ds\exp[-it_1\omega_1+i s\omega_2].$
All the integrals in the above expressions are easily calculated and the approximate evolution operator given by Eq.~\eqref{new} and \eqref{trot}.
 Thus, the perturbation expression for the evolution operator, given by the suggested method, is equal to

 \begin{equation}
\!\!\!\!\!\!\!\!\!\!\!\!\!\!\!\!\!\!\!\!\!\!\!\!\!U(t,t_0)= \left(
\begin{array}{ccc}
e^{f_2\left(t,t_0\right)} & -e^{f_1\left(t,t_0\right)}\Omega _1
   \Omega _2 g^*\left(t,t_0\right) &
   -i
  h_2^*\left(t,t_0\right) e^{f_1^*\left(t,t_0\right)+f^*_2\left(t,
   t_0\right)}\\
 -e^{f_2\left(t,t_0\right)}g\left(t,t_0\right) \Omega _1
   \Omega _2 & e^{f_1\left(t,t_0\right)}
   & -i
h_1^*\left(t,t_0\right) e^{f_1^*\left(t,t_0\right)+f^*_2\left(t,
   t_0\right)} \\
 -i e^{f_2\left(t,t_0\right)}h_2\left(t,t_0\right) & -i
   h_1\left(t,t_0\right)e^{f_1\left(t,t_0\right)} &
   e^{f_1^*\left(t,t_0\right)+f^*_2\left(t,
   t_0\right)} \\
\end{array}
\right),
 \end{equation}
    where
   \begin{equation}
    h_i(t,t_0)=\int_{t_0}^t\Omega_i(s)ds=-\frac{i \Omega _i \left(e^{i t
   \omega _i}-e^{i t_0 \omega
   _i}\right)}{\omega _i}.\label{tcl}
   \end{equation}
 
\begin{figure}
\begin{center}
\includegraphics[scale=0.8]{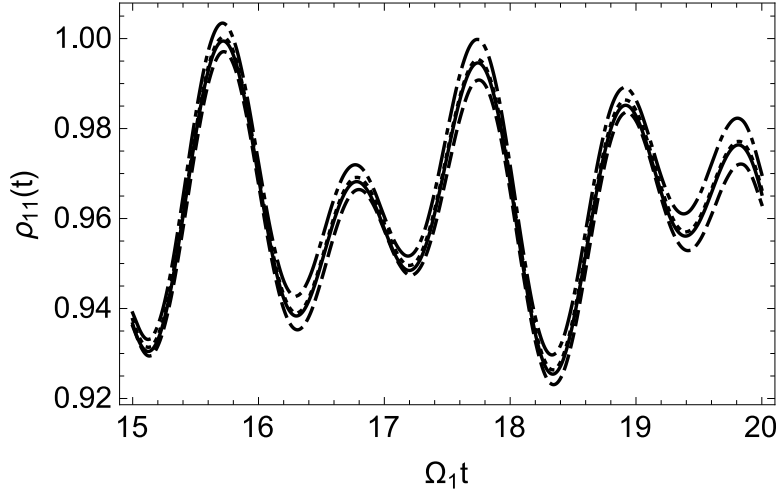}
\caption{The evolution of the ground state $\rho_{11}(t).$ Solid curve is the ``exact'' solution, dashed curve is the suggested approximation with the help of Eq.~\eqref{new}, dot-dashed curve is the standard second order approximation and dotted curve is average value of the standard and suggested approximations.
Parameters in the system are $\Omega_1=1,\, \Omega_{2}=0.7\Omega_1, \, \omega_1=1.3\Omega_1, \, \omega_2=5.3\Omega_1$.  The time step for approximation technique is $\Omega t=0.1$} \label{result}
\end{center}
\end{figure}

For comparison we also calculate the second order Dyson expansion $U(t,t_0)\approx I+i\int_{t_0}^tds H(s)-\int^t_{t_0}dt' \int_{t_0}^{t'}dsH(t')H(s).$ The explicit form is 
\begin{eqnarray}
\!\!\!\!\!\!\!\!\!\!\!\!\!\!\!\!\!\!\!\!\!\!\!\!\!U(t,t_0)= \left(
\begin{array}{ccc}
1+{f_2\left(t,t_0\right)} & -\Omega _1
   \Omega _2 g^*\left(t,t_0\right) &
   -i
  h_2^*\left(t,t_0\right) \\
 -g\left(t,t_0\right) \Omega _1
   \Omega _2 & 1+{f_1\left(t,t_0\right)}
   & -i
h_1^*\left(t,t_0\right) \\
 -i h_2\left(t,t_0\right) & -i
   h_1\left(t,t_0\right) &
  1+{f_1^*\left(t,t_0\right)+f^*_2\left(t,
   t_0\right)} \\
\end{array}
\right).\label{dys}
 \end{eqnarray}

 Even in this example one can clearly see that resulting expressions differ significantly.

The results of the calculation for the population of the ground state $|1\rangle$ from initial state  $|\psi(0)\rangle=|1\rangle,$ i.e. $|\langle 1|U(t,0)|1\rangle|^2,$ is presented in Fig.~\ref{result}. In the same figure we plot also the ``exact'' result following from the numerical solution of the Schr\"{o}dinger equation with Hamiltonian \eqref{ham1} and the result of the standard second order approximation of the evolution operator \eqref{dys} with iterative procedure \eqref{trot}. The time step of the iterative procedure was $\Omega_1 t=0.1$ and the Fig.~\ref{result} shows the result from 150 to 200 iterations. As one can see both the iterative procedures reproduce the exact dynamics quite accurately, even for such a large time step, but the perturbation formula Eq.~\eqref{new} works better, in this particular example. At the same time, the deviations from the exact solution have different signs for the iterative schemes. This fact allows to say that the approximation with the help of Eq.~\eqref{new} are not equivalent to the standard perturbation expansion and can be considered as an alternative to traditional approximation schemes. It is interesting, that the last fact allows to improve both the approximations just by averaging the resulting expressions. The average value of the two approximation schemes is also presented in Fig. 1 and one can see that the averaging really gives more accurate results.

\subsection{Partition function of a finite XY-spin chain.} In this paragraph we consider another interesting application of the above theory. Let us calculate the partition function $Z=\mathrm{Tr}\exp[-\beta H],$ where $\beta$ is the inverse temperature of a finite dimensional XY-spin chain. The Hamiltonian of the system is 
\begin{equation}\label{ham2}
H=A\sum_i^N (\sigma_+^i\sigma_-^{i+1}+\sigma_-^i\sigma_+^{i+1}),
\end{equation}
where A is the constant of the spin-spin interaction, $\sigma_\pm$ are the usual Pauli matrices and $N$ is number of sites in the chain. We also assume the periodic boundary conditions.

To apply the above theory let us notice that formally $\exp[-\beta H]=U(-i\beta),$ so, the calculation of the partition function is equivalent to the calculation of the evolution operator with pure imaginary time. Also, one has to keep only the diagonal part of the evolution operator. The diagonal part of the evolution operator gives the expression $\exp[\int_{t_0}^t \mathcal{K}(s)ds]$ and can be calculated with any accuracy.

The Hamiltonian \eqref{ham2} does not have non-zero diagonal elements. Thus, the second order approximation of the partition function is $\mathrm{Tr}\exp[\mathcal{P}\beta^2 H^2/2].$ Particularly, for the spin chain consisting of  10 sites the second order approximation gives
\begin{equation}
Z_{TCL}=8 \exp\left[\frac{5 (A\beta)^2}{2}\right] \cosh \left(\frac{(A\beta) ^2}{2}\right)\left(44 \cosh \left((A\beta)^2\right)+\cosh \left(2 (A\beta)^2\right)+83\right),\label{part1}
\end{equation}
while the traditional second order approximation $Z=\mathrm{Tr}(I+\beta^2/2 H^2)$ gives
\begin{equation}\label{part2}
Z_{Dyson}=1024 + 2560 (A\beta)^2.
\end{equation}
The comparison of the approximation results with the exact partition function is presented in Fig. 2. It is clearly seen that Eq.~\eqref{part1} reproduces the partition function for small $A\beta$ better than the corresponding result from Eq.~\eqref{part2}. The deviations of the both methods from the exact result again have different signs. We plot also the average value $1/2(Z_{TCL}+Z_{Dyson})$ in Fig. 2. The average value of the two methods improves the results.

%At the same time for $A\beta\approx 1$ Eq.~\eqref{part1} gives too fast growing and reproduce the partition function worse. 

%Notice, also, that as and for previous example the deviations from the exact partition function has a different sign for 

\begin{figure}
\begin{center}
\includegraphics[scale=0.8]{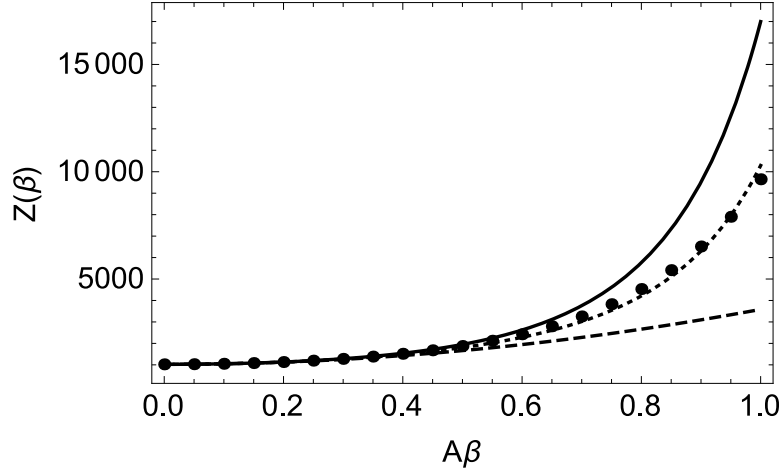}
\caption{The comparison of the suggested approximation Eq.~\eqref{part1} and the traditional approximation Eq.~\eqref{part2} with the exact partition function. The solid curve is Eq. \eqref{part1}; the dashed curve is Eq. \eqref{part2} and the dots are the exact partition function, the dotted curve is the average value of Eq. \eqref{part1} and Eq. \eqref{part2}.} \label{result2}
\end{center}
\end{figure}

\section{Conclusions.} 
In this paper, we study an evolution operator with the help of the projection operator technique. Applied to the Schr\"{o}dinger  equation the procedure, which is an analogue to the famous time-convolutionless projection technique in the theory of open quantum systems and statistical physics, leads to the alternative expression \eqref{new} for the evolution operator. This expression can be considered as the starting point for a systematic perturbative investigation of the evolution operator. The resulting perturbative expression differs from the one derived by cutting off the chronological exponent and any other perturbation expansions. The approximated evolution operator can be used to recover the full evolution operator in the spirit of the Lie-Trotter formula through the iterative procedure \eqref{iter}. The iterative procedure can be considered as a numerical scheme for the simulation of a wide class of linear differential systems. The deviation of the iterative scheme from the exact result seems to have a different sign in comparison with the using of the truncated chronological exponent and simple averaging of the two approximations improves the results. The interest- ing feature of the suggested technique is the possibility to study independently the diagonal elements of the evolution operator, which is given by the solution of Eq. \eqref{rel2}. This feature may be very useful for the calculation of the partition function $Z=\mathrm{Tr}\exp[-\beta H],$  which is formally derived by substitution  $t\rightarrow -i\beta$ for the evolution operator.

\section*{Acknowledgements}
This work is based upon research supported by the South African
Research Chair Initiative of the Department of Science and
Technology and National Research Foundation.

\end{document}